\documentclass[twocolumn]{aastex62}

\usepackage{lineno}
\usepackage{longtable}

\def\kms{km~s$^{-1}$}

\newcommand{\xit}{\ensuremath{\xi _{t}}}

\newcommand{\vsini}{\ensuremath{v_{{\mathrm e}}\sin i}}
\newcommand{\teff}{\ensuremath{T_\mathrm{eff}}}

\newcommand{\logg}{\ensuremath{\mathrm{log}\ g}}

\graphicspath{{./}{figures/}}

\received{July 4, 2022}
\revised{June 13, 2023}
\accepted{June 14, 2023}
\submitjournal{AJ}
\shorttitle{Variability of $\alpha$ Sex}
\shortauthors{Monier et al.}

\begin{document}
	
\title{The unexpected optical and ultraviolet variability of the standard star $\alpha$ Sex (HD~87887)}
\correspondingauthor{Richard Monier}
\email{Richard.Monier@obspm.fr}

\author[0000-0002-0735-5512]{Richard Monier}
\affiliation{LESIA, UMR 8109, Observatoire de Paris et Universit\'e Pierre et Marie Curie Sorbonne Universit\'es, place J. Janssen, Meudon, France.}

\author[0000-0001-7402-3852]{Dominic M. Bowman}
\affiliation{Institute of Astronomy, KU Leuven, Celestijnenlaan 200D, B-3001 Leuven, Belgium}

\author[0000-0002-4834-2144]{Yveline Lebreton}
\affiliation{LESIA, UMR 8109, Observatoire de Paris et Universit\'e Pierre et Marie Curie Sorbonne Universit\'es, place J. Janssen, Meudon, France.}
\affiliation{Univ Rennes, CNRS, IPR (Institut de Physique de Rennes) - UMR 6251, F-35000 Rennes, France.}

\author{Morgan Deal}
\affiliation{Instituto de Astrof\'isica e Ci\^encias do EspaÃ§o, Universidade do Porto, CAUP, Rua das Estrelas, PT4150-762 Porto, Portugal}
\affiliation{LESIA, UMR 8109, Observatoire de Paris et Universit\'e Pierre et Marie Curie Sorbonne Universit\'es, place J. Janssen, Meudon, France.}

\begin{abstract}
The analysis of the available TESS light curves of $\alpha$ Sex (HD~87887) reveals low-frequency pulsations with a period of about 9.1 hours in this spectroscopic A0~III standard star. The IUE observations in December 1992 reveal large flux variations both in the far UV and in the mid UV which are accompanied by variations of the brightness in the V band recorded by the Fine Error Sensor on board IUE. The ultraviolet variability could be due to an eclipse by an hitherto undetected companion of smaller radius, possibly 2.5 $R_\odot$ but this needs confirmation by further monitoring possibly with TESS.
An abundance determination yields solar abundances for most elements. Only carbon and strontium are underabundant and titanium, vanadium and barium mildly overabundant.
 Identification is provided for most of the lines absorbing more than 2\% in the optical spectrum of $\alpha$ Sex. Stellar evolution modeling shows that $\alpha$ Sex is near the terminal-age main sequence, and its mass, radius and age are estimated to be $M=2.57\pm0.32\, M_\odot$, $R=3.07\pm0.90\, R_\odot$, $A =385\pm77\, \mathrm{Myr}$, respectively.
\end{abstract}


\section{Introduction} \label{sec:intro}

Although $\alpha$ Sex is a bright and standard A0~III giant \citep{1987ApJS...65..581G}, it has not been extensively studied for a star of its brightness: only 91 references can be found in SIMBAD\footnote{http://simbad.u-strasbg.fr/simbad/}.
The most recent abundance analysis is that of \citet{2003A&A...406..987P} who used optical spectra and derived abundances for 19 elements. They found mostly solar abundances including helium. The only species that deviated from solar abundances are scandium, which is underabundant, sulfur and calcium marginally underabundant, manganese marginally overabundant and barium overabundant.
A new determination of the iron abundance was made by \cite{2014PASP..126..505A} who found a slight underabundance. 

Among intermediate-mass main-sequence stars of spectral type A and F, the most common type of pulsator are the $\delta$~Sct stars. These stars have low-radial order pressure modes with periods of order of hours that are excited by a heat-engine mechanism \citep{Breger2000}. Approximately 50\% of main-sequence A and early-F stars are $\delta$~Sct stars based on modern high-precision space photometry \citep{Bowman2018, Murphy2019}. The identification of modeling of stellar pulsations, known as asteroseismology \citep{Aerts2010}, yields important constraints on the physical processes at work within stars such as rotation, mixing and atomic diffusion. Therefore, the identification of high-quality pulsating stars is essential for follow-up modelling.

In this paper, we report on work based on new high-precision light curves from the NASA Transiting Exoplanet Survey Satellite (TESS) mission \citep{Ricker2015} and on a new abundance analysis. The TESS light curve reveals that $\alpha$ Sex is a variable star with multi-periodic pulsations with periods of the order of several hours. The IUE archival observations of $\alpha$ Sex over 3 days in December 1992 are also analysed

This article is organised as follows. Section 2 describes the TESS light curves of $\alpha$ Sex and their analysis, section 3 the analysis of the IUE spectra,
and section 4 the abundance determinations for 19 chemical elements, in section 5, we use the SPInS stellar parameter inference program \citep{Lebreton2020} and the BaSTI-IAC grid of stellar models \citep{Hidalgo2018}
to derive the evolutionary status, mass, radius, and age of $\alpha$ Sex. We discuss the nature of $\alpha$ Sex and conclude in the final section.

\section{The TESS light curves of $\alpha$ Sex and their analysis} \label{sec:tess}

The TESS mission observed $\alpha$~Sex in sectors 8, 35 and 45 in its short cadence (i.e. 2-min) mode. We retrieved both the simple aperture photometry (SAP) and pre-data search conditioning (PDC-SAP) 2-min light curves from the Mikulski Archive for Space Telescopes (MAST\footnote{\url{https://archive.stsci.edu/}}), which are extracted from the target pixel files using NASA's SPOC pipeline (see \citealt{Jenkins2016b} for details). Since $\alpha$~Sex is a relatively bright star for the TESS mission, it is moderately saturated in its target pixel files. However, the aperture mask assigned by the SPOC pipeline includes sufficient pixels, which includes the short bleed columns, to extract a light curve. In such cases, TESS light curves are more than adequate at detecting and characterising pulsating stars (see e.g. \citealt{Bowman2022}). We checked for possible sources of contamination, but could not verify any known and sufficiently bright background or nearby sources. Using the {\sc Lightkurve} software \citep{Lightkurve2018} software in combination with Gaia astrometry \citep{DR3}, there are a only a few very faint (Gaia $G > 14$~mag) sources located within the assigned target pixel aperture mask, hence their flux contribution is negligible.

\begin{figure*}
\center
\includegraphics[width=0.99\textwidth]{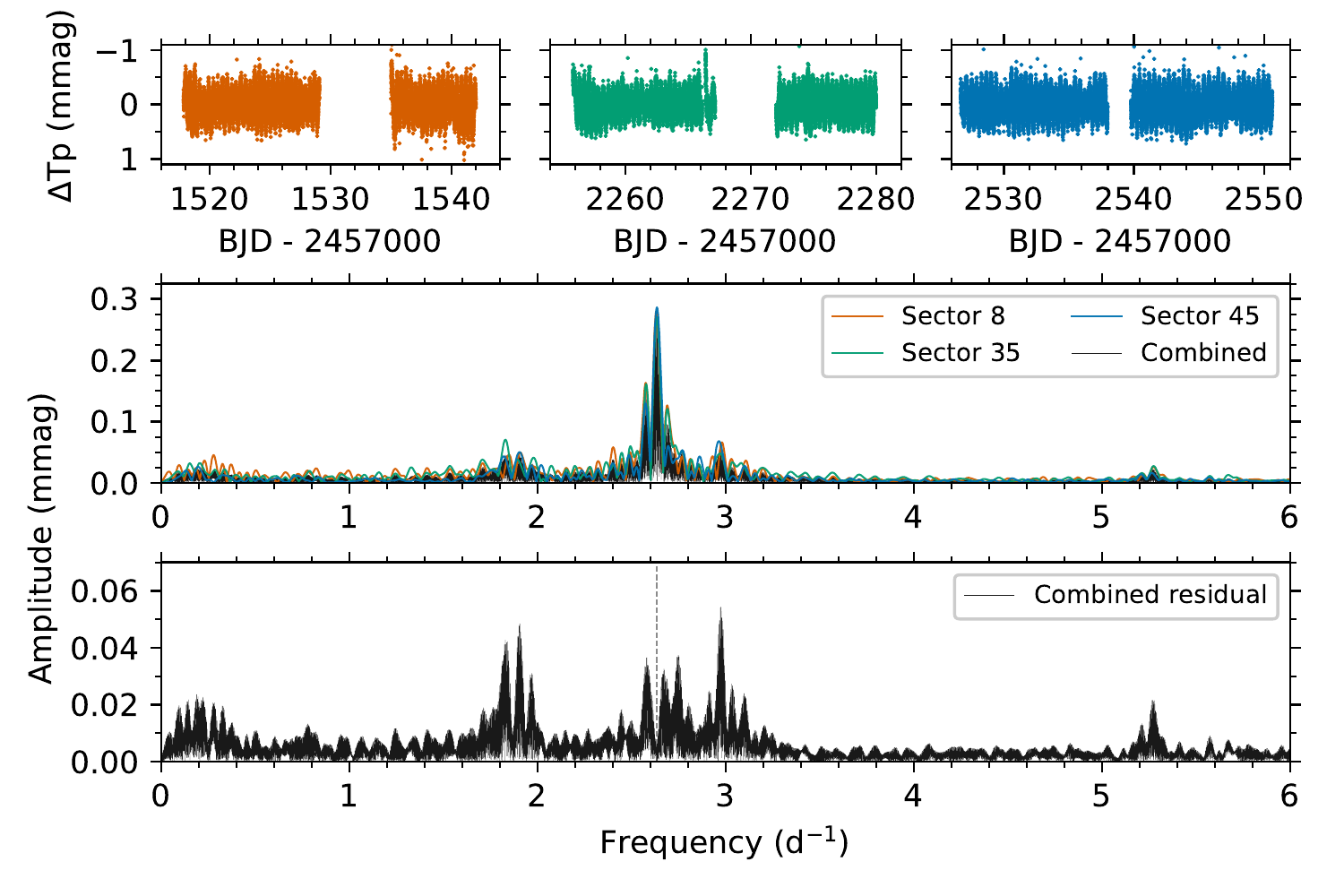}
\caption{{\it Top panels:} TESS light curves for sectors 8, 35 and 45. {\it Middle panel:} Amplitude spectra of each TESS sector and all three sectors combined. {\it Bottom panel:} Residual amplitude spectrum of all three combined sectors after the dominant frequency has been removed.}
\label{fig:TESS}
\end{figure*}

Given the large gaps between the sectors 8, 35 and 45 light curves, we opted to analyse them separately for signatures of pulsations to avoid issues arising from the complex spectral window pattern in a combined light curve. Each TESS sector light curve maximally spans approximately 24~d with varying duty cycles depending on the specific sector. This yields a resultant frequency resolution following the Rayleigh criterion of approximately 0.042~d$^{-1}$ for an individual sector. We converted the extracted PDC-SAP 2-min light curves to have units of magnitudes and show them in Fig.~\ref{fig:TESS}. We calculated discrete Fourier transforms \citep{Kurtz1985} and show the resultant amplitude spectra for each sector and the combined sectors 8, 35 and 45 light curve in the middle panel of Fig.~\ref{fig:TESS}. A dominant frequency of 2.63~d$^{-1}$, corresponding to a period of 9.1~hr, is apparent in the amplitude spectra of all three individual light curves. Additional multi-periodic variability is present in the frequency range of 1.8~d$^{-1}$ up to 5.3~d$^{-1}$, with amplitudes ranging up 0.3~mmag. 

We performed iterative pre-whitening to extract significant frequencies for each of the three individual TESS sectors. Significant frequencies are those that have an amplitude signal-to-noise ratio (S/N) of $\geq 5$, in which the noise is calculated using a symmetric local window centred at the location of the extracted frequency in the residual amplitude spectrum \citep{Bowman2021}. In our frequency analysis of the individual TESS sectors, two significant frequencies are extracted: the dominant frequency and a second indistinguishable for its harmonic given the low resolving power of a single TESS sector. Specifically in sector 45, however, several additional frequencies are detected within the frequency regime of the dominant frequency. This is not the signature of rotational modulation, but in fact is evidence of multi-periodic pulsations. Moreover, for such a frequency to be caused by rotational modulation, this implies an extremely rapid rate of surface rotation, which we deem unlikely given the known projected surface rotation rate from spectroscopy $\vsini$ = 21 \kms \citep{abt02}. 

We provide the list of significant frequencies for each sector, which were optimised using a multi-frequency non-linear least-squares fit to the light curve \citep{Bowman2021}, in Table~\ref{table:freqs}. In the bottom panel of Fig.~\ref{fig:TESS}, we show the residual amplitude spectrum after the dominant frequency has been optimised and removed from the combined sectors 8, 35 and 45 light curve. This clearly demonstrates the multi-periodic variability of $\alpha$~Sex between 1.8 and 5.3~d$^{-1}$, which is affected by the complex spectral window pattern resulting from the combined light curves. 

\begin{table}
\caption{Frequencies, amplitudes and phases of the significant pulsation modes in $\alpha$~Sex, and their $1\sigma$ uncertainties calculated from a non-linear least-squares fit to the light curve.}\label{table:freqs}
\begin{center}
\begin{tabular}{r r r}
\hline
\hline
Frequency & Amplitude & Phase \\
(d$^{-1}$) & (mmag) & (rad) \\
\hline
Sector 8 \\
$2.6313 \pm 0.0001$	&	$0.280 \pm 0.002$	&	$0.221 \pm 0.007$	\\
$5.2745 \pm 0.0017$	&	$0.024 \pm 0.002$	&	$-2.727 \pm 0.083$	\\
\hline
Sector 35 \\
$2.6333 \pm 0.0001$	&	$0.271 \pm 0.002$	&	$1.157 \pm 0.007$	\\
$5.2753 \pm 0.0013$	&	$0.028 \pm 0.002$	&	$-0.863 \pm 0.063$	\\
\hline
Sector 45 \\
$1.8139 \pm 0.0008$	&	$0.0409 \pm 0.002$	&	$-1.042 \pm 0.034$	\\
$1.9054 \pm 0.0006$	&	$0.0537 \pm 0.002$	&	$2.279 \pm 0.026$	\\
$2.5791 \pm 0.0009$	&	$0.0612 \pm 0.002$	&	$-2.150 \pm 0.031$	\\
$2.6329 \pm 0.0004$	&	$0.2800 \pm 0.002$	&	$2.290 \pm 0.010$	\\
$2.6826 \pm 0.0042$	&	$0.0316 \pm 0.002$	&	$-3.017 \pm 0.081$	\\
$2.7382 \pm 0.0019$	&	$0.0407 \pm 0.002$	&	$0.714 \pm 0.041$	\\
$2.9678 \pm 0.0005$	&	$0.0621 \pm 0.002$	&	$1.156 \pm 0.023$	\\
$3.0987 \pm 0.0014$	&	$0.0231 \pm 0.002$	&	$3.046 \pm 0.062$	\\
$5.2634 \pm 0.0018$	&	$0.0167 \pm 0.002$	&	$1.347 \pm 0.082$	\\
\hline
\end{tabular}
\end{center}
\end{table}

\section{The ultraviolet variability of $\alpha$ Sex}\label{sec:iue}

 Nine high resolution SWP and LWP spectra of $\alpha$ Sex were obtained with the International Ultraviolet Explorer from 24 to 27 December 1992 through the large apertures in the frame of program  NA026 (PI: Richard Monier). These spectra are calibrated into absolute fluxes, their resolving power is about 25000 and their signal-to-noise ratios is typically a 10-20. They were retrieved from the MAST, and they are collected in Table~\ref{tab:obs}.
\begin{table*}
\caption{Log of IUE OBSERVATIONS}
\label{tab:obs}
\centering
\begin{tabular}{||c|c|c|c|c|c|}
\hline \hline
Spectrum      & Observation& Resolution & Observation     & Exposition  &   FES\\ 
             & Date       &            &  time      & Time [s]        &   counts  \\ \hline
SWP 46576 & 1992-12-24 & High     & 13:47:20  & 390  & 3587  \\
LWP 24572 & 1992-12-24 & High     & 13:59:05  & 180  & 3696 \\
SWP 46577 & 1992-12-24 & High     & 16:12:40  & 1200 & 3632\\
LWP 24573 & 1992-12-24 & High     & 16:27:19  & 390  & 3841 \\
\hline
SWP 46584 & 1992-12-25 & High     & 15:56:55  & 1200 & 3321\\
LWP 24589 & 1992-12-25 & High     & 16:25:57  & 4200  & 3227\\
\hline
SWP 46598 & 1992-12-27 & High     & 10:14:56  & 1800 & 5407\\	     
LWP 24605 & 1992-12-27 & Low     & 11:05:42   & 420 & 5542 \\
LWP 24606 & 1992-12-27 & High     & 12:50:53 & 420 & 5443\\
SWP 46599 & 1992-12-27 & High     & 16:46:54  & 600 & 5401\\
\hline \hline
\end{tabular}
\end{table*}

\begin{figure*}
\plottwo{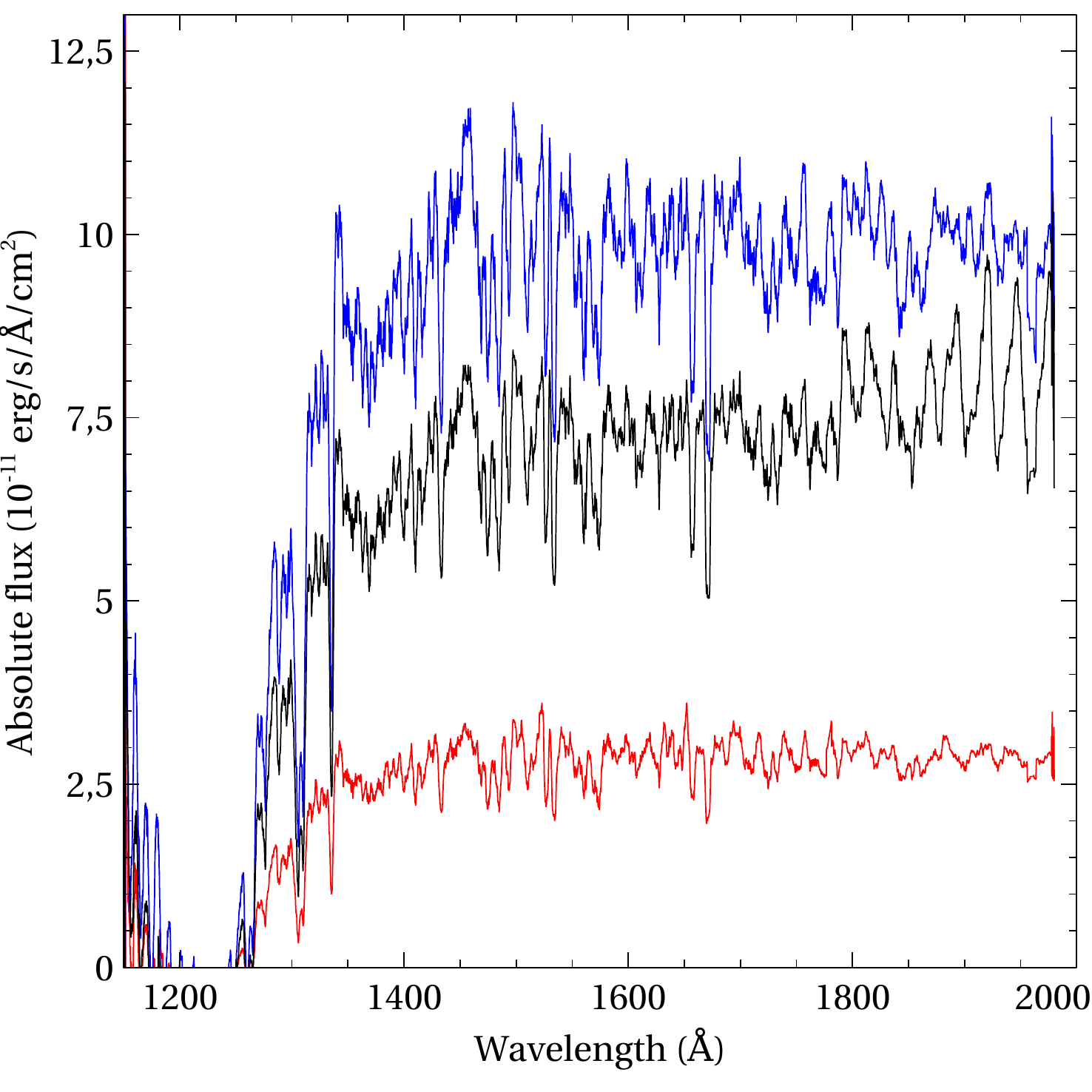}{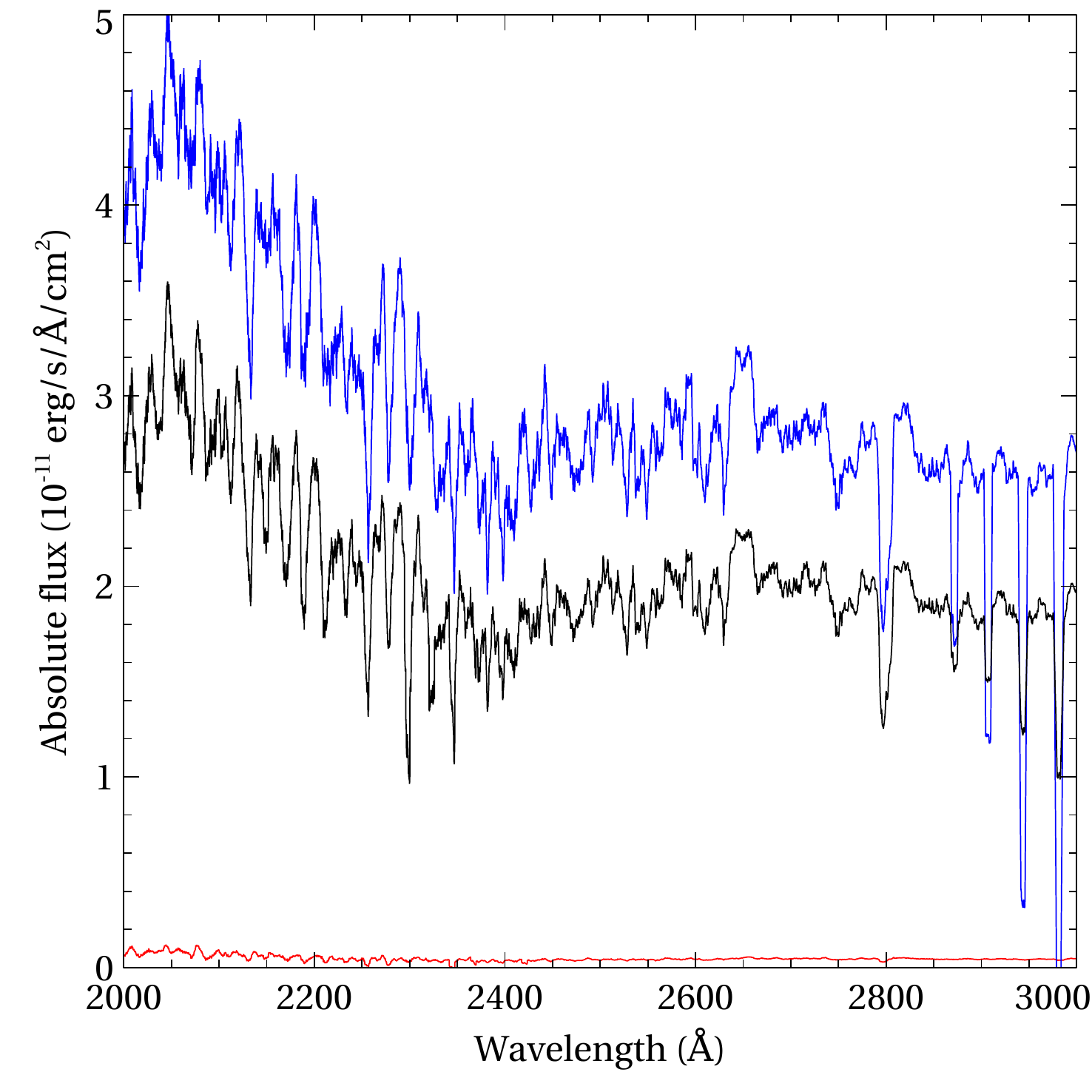}
\caption{Left: The far UV flux variations of $\alpha$ Sex (SWP~46599, FUV max in blue; SWP~46577 FUV min in red), Right: The mid UV flux variations (LWP~24606: midUV max in blue; LWP2~4589: midUV min in red).}
\label{fig:2}
\end{figure*}

These spectra have been degraded to a lower resolution of about 7 \AA\ comparable to the IUE low resolution in order to highlight the variations. The large variations in the far and mean ultraviolet are shown in Figure \ref{fig:2}. In the far UV, four pseudo continuuum windows (regions free of strong lines) are present at 1284, 1342, 1457 and 1756 \AA. The ratio of maximum (SWP~46599) to minimum flux (SWP~46577) is about the same in these three windows: 1.35 $\pm$ 0.03.
In the mean UV, two windows are observed at 2047 and 2650 \AA, where the ratio is about 1.38 $\pm$ 0.01, discarding the spectrum with very low flux LWP~24589. Including this spectrum, the ratio becomes 44 at 2047 \AA\ and 57 at 2650 \AA.
Before each of these observations, the brightness of $\alpha$ Sex was monitored  with the Fine Error Sensor in a broad spectral band near 5000 \AA. The FES counts do folllow the UV flux variations: they are larger on December 27 (around 5450 counts) , minimum on December 25 (around 3270 counts) and intermediate on December 24 (around 3690 counts) which means that the UV flux varied in phase with the brightness of the star at 5000 \AA. At ultraviolet maximum, the lines in SWP~46599  are consistently redshifted by about 13.7 \kms with respect to the spectrum at FUV minimum SWP~46577.

The large diming of the flux by about 70 \% at minimum in the far UV compared to the maximum flux could be due to a partial eclipse of $\alpha$ Sex by an
hitherto undetected companion. The duration of the eclipse and the shape of the light curve are poorly constrained with the available data, it is therefore difficult to derive information on the radii, temperatures and masses. 
We can crudely estimate the radius of the secondary star by using the relationship between the flux decrease:
\begin{equation}
\frac{f_{\rm min}}{f_{\rm max}} = \left(\frac{R_{B}}{R_{A}}\right)^2
\end{equation}
and the radii $R_{A}$ of $\alpha$ Sex and $R_{B}$ of the putative companion. Assuming a radius of about 3.0 $R_{\odot}$ for $R_{A}$ (see section 3), this yields a radius $R_{B}$ close to 2.5 $R_{\odot}$.
If the system is indeed eclipsing, it is seen edge-on ($i=\frac{\pi}{2}$). We can estimate a rotation period of 7 days and 19 hours by using the projected equatorial velocity and the estimated radius in section 6 where we discuss the evolutionary status of $\alpha$ Sex. Since no eclipse is seen in the current TESS data, we can place a lower limit on the orbital period of about 28 days, which means that the semi-major axis of the ellipse must be large. \cite{Eggleton2008} mentioned that $\alpha$ Sex could be a binary, presumably with a long period. \cite{Kervella2019} also think that $\alpha$ Sex may be a binary system considering the renormalised unit weight error from Gaia DR2. Note that the large ultraviolet variations we observe for $\alpha$ Sex do not resemble those of $\delta$ Scuti observed by \cite{Monier1991} throughout its pulsation cycle. The ultraviolet variations of $\delta$ Scuti have modest amplitudes which increase towards shorter wavelengths as expected for a change in effective temperature during the pulsation cycle. This is not the trend we observe for $\alpha$ Sex for which the amplitude of the flux variations does not increase towards shorter wavelengths (the far and mid-UV fluxes vary by a similar amount with time).

\section{The abundance pattern of $\alpha$ Sex} \label{sec:iue}

\subsection{Observations}

Four I  high resolution profiles (R = 65000) of $\alpha$ Sex have been fetched from the Polarbase archive\footnote{http://polarbase.irap.omp.eu/}. These profiles were acquired on 10 May 2018   with the spectropolarimeter NARVAL \citep{Petit} installed at the 2 meter TBL at Pic du Midi Observatory.
NARVAL is a cross-dispersed \'echelle spectrograph mounted on a bench and fed with a fiber from a Cassegrain-mounted polarimeter unit with a wavelength coverage of 3690 up to 10480 \AA.
The individual I profiles which have a signal-to-noise ratio of 200 around 5000 \AA\ were coadded into a mean spectrum of signal-to-noise of 350. This final spectrum has been sliced into 200 \AA\ wide intervals which were then normalised to a continuum by fitting a cubic spline through narrow regions free of absorption lines.

\subsection{Fundamental parameters}

The effective temperature (\teff) and surface gravity (\logg) of $\alpha$~Sex were determined using the UVBYBETA code developed by \cite{Napiwotzki}. This code is based on the \cite{1985MNRAS.217..305M} grid, which calibrates the $uvby\beta$ photometry in terms of \teff \ and \logg. The photometric data was taken from \cite{1998A&AS..129..431H}.
The derived effective temperature is \teff = 9950  $\pm$125~K and \logg = 3.60 $\pm$0.25 dex, respectively (see Sec. 4.2 in \citealt{Napiwotzki}).
This value is in good agreement with previous determinations: \cite{2014PASP..126..505A} derived \teff = 9875 K from spectrophotometry and the fit to the $H_{\gamma}$ line; \cite{2012MNRAS.427..343M} derived \teff = 9984 K by comparing the spectral energy distribution of $\alpha$~Sex to model atmospheres; and
\cite{2003A&A...406..987P} derived \teff = 9950 K from calibration of uvby$\beta$ photometry.

\subsection{Abundance determination}

\subsubsection{Model atmospheres and spectrum synthesis}

The ATLAS9 code \citep{Kurucz} was used to compute a first model atmosphere for the effective temperature and surface gravity of $\alpha$ Sex assuming
a plane parallel geometry, a gas in hydrostatic and radiative equilibrium and local thermodynamical equilibrium. The ATLAS9 model atmosphere contains 72 layers with a regular increase in $\log \tau_{\rm Ross} = 0.125$ and was calculated assuming a solar chemical composition \citep{1998SSRv...85..161G}. It was converged up to $\log \tau = -5.00$ in order to attempt reproduce the cores of the Balmer lines.
This ATLAS9 version uses the new opacity distribution function (ODF) of \cite{Castelli2003} computed for that solar chemical composition. Once a first set of elemental abundances was derived using the ATLAS9 model atmosphere, the atmospheric structure was recomputed for these abundances using the Opacity sampling ATLAS12 code \citep{atlas12,atlas12_2}. New slightly different abundances were then derived and a new ATLAS12 model recomputed until the abundances of iteration (n-1) differed of those of iteration (n) by less than $\pm$ 0.10 dex.

The abundances of nineteen chemical elements have been derived by iteratively
adjusting synthetic spectra to the normalized spectra and looking for the best fit to carefully selected unblended lines. Specifically, synthetic spectra were computed assuming LTE using \cite{Hubeny92} SYNSPEC49 code which computes lines for elements up to Z=99. The synthetic spectra were further convolved with a rotation parabolic profile for $\vsini$ = 21 \kms \citep{abt02} and the appropriate FWHM of the instrumental profile of NARVAL. The projected equatorial velocity has been checked by modeling the \ion{Fe}{2} lines in the range 4500-4550 \AA, they all yield a 
$\vsini$ of about  20.0 $\pm$ 1.0 \kms which agrees well with the value provided by \cite{Royer2002}.
In order to derive the microturbulenct velocity of $\alpha$ Sex, we simultaneously derived the iron abundance [Fe/H] for 50 unblended \ion{Fe}{2} lines and a set of microturbulent velocities ranging from 0.0 to 2.0 \kms. The adopted microturbulent velocity, $\xit = 1.5$ \kms, minimizes the standard deviations, i.e. for that value all \ion{Fe}{2} lines yield a similar iron abundance.





We used only unblended lines to derive the abundances. A grid of synthetic spectra was computed with SYNSPEC49 \citep{Hubeny92} to model each selected unblended line of the nineteen elements  for $\alpha$ Sex. For each modeled transition, the adopted abundance is that which provided the best fit calculated with SYNSPEC49 to the observed normalized profile. Computations were iterated varying the unknown abundance until minimization of the $\chi^2$ between the observed and synthetic spectrum was achieved over the spectral range limited to $\pm$ 1.5 \AA\ from the line center. The selected lines are well separated from their neighbours allowing to place the continuum properly on both wings of the line. For a given element, the final abundance is a weighted mean of the abundances derived for each transition (the weights are derived from the NIST grade assigned to that particular transition).

\section{Abundance determinations and line identifications for $\alpha$ Sex} \label{sec:abundances}

The determined abundances for $\alpha$ Sex, expressed relative to hydrogen as $\log(n_{X}) - 12.0$ (adopting $\log(n_{H}) = 12.0$) and their errors (standard deviations) are collected in Table \ref{tab:abund}. Solar abundances are taken from \cite{2007SSRv..130..105G}. We find that the abundances of $\alpha$ Sex are close to the solar composition.

\begin{figure}
\center
\includegraphics[width=\columnwidth]{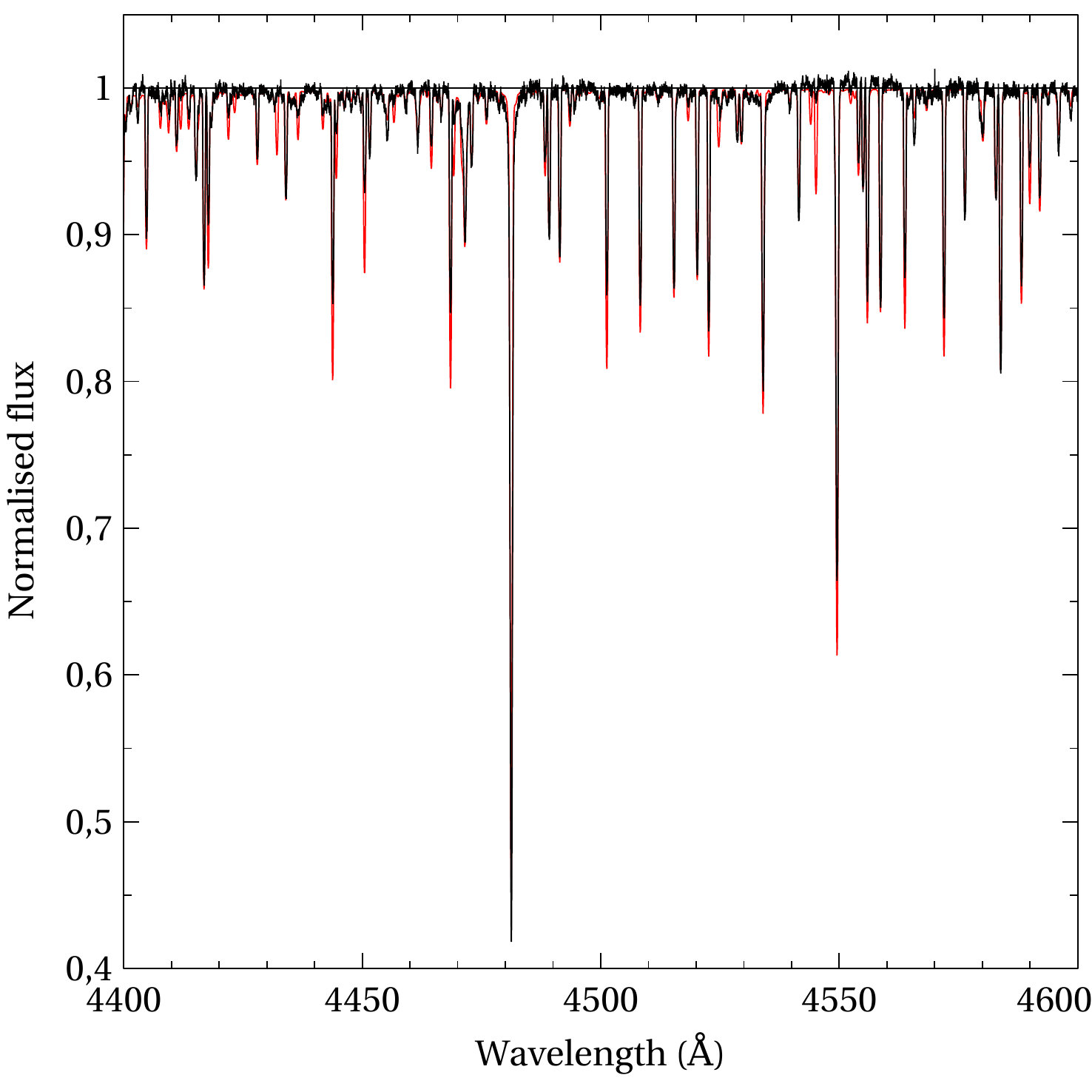}
\caption{Comparison of the observed mean spectrum of HD~87887 (in black) and the synthetic spectrum (in red) in the range 4500-4550 \AA}
\label{fig:spectrum}
\end{figure}

Helium, nitrogen, oxygen, magnesium, aluminium, silicon, phosphorus, sulfur, calcium, scandium, chromium, manganese, iron and nickel have solar abundances.
Only carbon and strontium are underabundant. Titanium, vanadium and baryum are mildly overabundant.
The final synthetic spectrum allows to identify most of the lines which absorb more than 2\% of the local continuum. It is compared to the observed spectrum in the range 4500 to 4550 \AA\ in Fig. \ref{fig:spectrum}.
The identifications of these lines are collected in Table \ref{tab:ident} where $E_{\rm low}$ is the energy of the lower excitation level involved in the transition.

\begin{table}
\caption{Elemental abundances and their errors for $\alpha$ Sex}
\label{tab:abund}
\centering
\begin{tabular}{cccc}
\hline  
 Element & Solar abundance & Absolute abundance & Error \\
  \hline
He & -1.07 & -1.07 & 0.32\\
 C  & -3.61 & -3.37 & 0.09\\
N  &  -4.22 & -4.39 & 0.20 \\
 O  & -3.34 & -3.31 & 0.19 \\
 Mg & -4.47  & -4.50 & 0.08\\
Al & -5.63 & -5.37 & 0.20\\ 
Si &  -4.49 & -4.49 & 0.18\\ 
 P  & -6.64 & -6.64 & 0.16\\
 S  &  -4.86 & -4.86 & 0.11\\
 Ca & -5.69  & -5.71 & 0.05 \\
 Sc & -8.83 &  -9.10 & 0.28\\
 Ti & -7.10 & -6.80 & 0.13\\
 V  & -8.00 & -7.60 & 0.08\\
 Cr & -6.36 & -6.36 & 0.05\\ 
 Mn & -6.61 & -6.57 & 0.11\\ 
 Fe & -4.55 & -4.69 & 0.10\\
 Ni & -5.97  &  -5.99 & 0.13 \\
 Sr & -9.08  & -9.23 & 0.09\\
 Ba & -9.83   & -9.34 & 0.19\\
\hline 
\end{tabular}
\end{table}

\section{The evolutionary status of $\alpha$ Sex} \label{sec:evolution}


\subsection{Estimations of mass, radius, and evolutionary stage} 

To estimate the mass, radius, and age of the star, we used the SPInS stellar model optimization tool \citep{Lebreton2020}. 
SPInS uses a Bayesian approach to find the probability distribution function of stellar parameters from a set of constraints. At the heart of the code is a Markov Chain Monte Carlo sampler coupled with interpolation within a pre-computed stellar model grid. Here, we used the BaSTI-IAC grid of stellar models \citep{Hidalgo2018}. This grid is for a solar-scaled heavy element distribution with the solar mixture taken from \citet{Caffau2011} complemented by \citet{Lodders2010}, which corresponds to $(Z/X)_\odot=0.0209$. The grid considers convective core overshooting included as an instantaneous mixing between Schwarzschild's convective limit up to layers at a distance $\alpha_\mathrm{ov}=0.2 H_P$ from it, where $H_P$ is the pressure scale height at the Schwarzschild limit. Microscopic diffusion is not taken into account. 
{In the calculation process, SPInS can incorporate various priors on the initial mass function (IMF), stellar formation rate (SFR) and metallicity distribution function (MDF). Here we took \citet{Kroupa2013} as IMF and no priors on the SFR and MDF.}
 


\begin{figure}
\center
\includegraphics[width=\columnwidth]{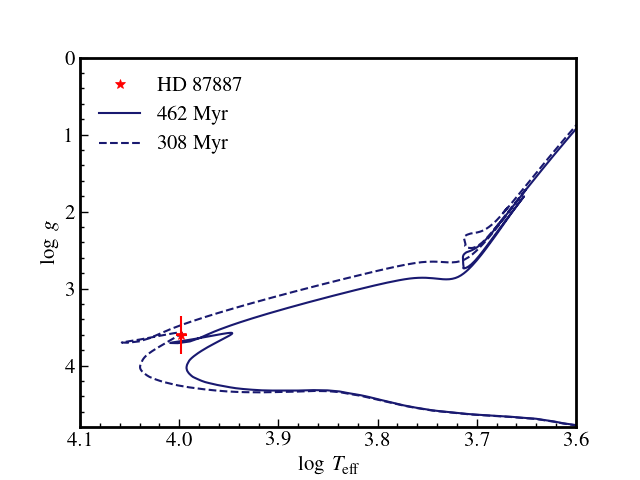}
\caption{Kiel diagram showing the observed position of HD 87887 together with two isochrones computed for the $\pm 1\sigma$ age-values inferred by SPInS.}
\label{fig:Kiel}
\end{figure}

With the observational constraints for HD 87887 derived in Sects.~\ref{sec:iue} and  \ref{sec:abundances} ($T_\mathrm{eff}= 9950 \pm 125\, \mathrm{K}, \log g= 3.60 \pm 0.25$ dex and $\mathrm{[Fe/H]}= -0.14 \pm 0.10$), SPInS provided as mean values an age $A=385\pm77\, \mathrm{Myr}$, a mass $M=2.57\pm0.32\, M_\odot$, a radius $R=3.07\pm0.90\, R_\odot$, a mean density $\overline{\rho} \simeq 0.13\pm 0.11\, \mathrm{g\,cm^{-3}}$, and a luminosity $L=90\pm52\, L_\odot$. The observed  position of the star in the Kiel diagram is provided in Fig. \ref{fig:Kiel} together with the isochrones for the $\pm 1\sigma$ age-values inferred by SPInS showing that the star evolves at the vicinity of the main sequence turn-off. 
We can compare the inferred radius with the observed one using the \citet{Swihart2017}'s angular diameter obtained by interferometry, $\theta\simeq 0.361\pm 0.065$ mas, and parallax values from the literature, that is $\varpi_\mathrm{G}=7.66 \pm 0.37$ mas \citep[Gaia DR3]{DR3} and $\varpi_\mathrm{H}=11.51\pm 0.98$ mas \citep[Hipparcos]{leeuwen07}. This leads to linear radius values of $R_\mathrm{G}\simeq 5.1\pm 1.2\, R_\odot$ and $R_\mathrm{H}\simeq 3.4\pm 0.9\, R_\odot$, respectively, such that $R_\mathrm{H}$ is closer to our inferred radius rather than $R_\mathrm{G}$. 
To assess our results, we also attempted to derive the stellar parameters using absolute magnitude and a color index in SPInS instead of the $T_\mathrm{eff}$ and $\log g$ values derived previously. From the Hipparcos and Tycho-2 constraints on $\varpi_\mathrm{H}, V_T=4.474\pm0.003$ mag and $(B-V)_T=-0.030\pm0.004$ mag \citep{Hog2000}, we found an age $A=350\pm25\, \mathrm{Myr}$, a mass $M=2.87\pm0.12\, M_\odot$, and a radius $R=3.66\pm0.34\, R_\odot$, all compatible with our results. On the other hand, we could not find a solution when using the absolute Gaia G-magnitude and (BP-RP) color index, which may indicate difficulties with the Gaia parallax. Indeed, the Gaia-DR3 parallax of $\alpha$ Sex should be taken with care for three reasons: (i) the star is bright; (ii) the astrometric excess noise of the source is around $2$ mas\footnote{see  \url{https://gea.esac.esa.int/archive/documentation/GDR2/Gaia\_archive/chap\_datamodel/sec\_dm\_main\_tables/ssec\_dm\_gaia\_source.html}
and \url{https://www.cosmos.esa.int/web/gaia/science-performance} (F. Arenou, private communication).}; and (iii) it is a member of a binary system  \citep{Kervella2019}.


\section{Discussion and Conclusions} \label{sec:discussion}

To place the detected variability caused by pulsations in an evolutionary context, we calculate the expected period of the fundamental radial pressure mode of $\alpha$~Sex using its derived parameters. Following \citet{Ledoux1945}, we estimate the period of the fundamental radial pressure mode of $\alpha$~Sex to be:
\begin{equation}
 P = \left(\frac{3\pi}{G \overline{\rho}} \times \frac{1}{3\Gamma_1-4}\right)^{\frac{1}{2}} ~ ,
\end{equation}
assuming the adiabatic exponent is constant throughout the star. For an ideal monoatomic gas for which  $\Gamma_1\approx 5/3$, this yields $\nu = 2.6\pm1.1$~d$^{-1}$ (i.e. $P = 9.3\pm4.1$ hours). This estimate is analogous and compatible with that of the empirical $Q = P \sqrt{\frac{\rho}{\rho_{\odot}}}$ relation for radial pressure modes in $\delta$~Sct stars \citep{Breger1975}, in which $Q=0.033$ for fundamental radial pressure modes. Our estimates are entirely consistent with the observed dominant frequency in the TESS light curve of $\alpha$~Sex. Therefore, we conclude that the observed variability is low-radial order p-mode pulsations. Given its advanced evolutionary stage, it is possible that the observed pulsation modes are of mixed pressure- and gravity-mode character \citep{Aerts2010}. However, future forward asteroseismic data based on much longer light curves with higher duty cycle are needed to confirm this.

Although the pulsations observed in the TESS light curve of $\alpha$ Sex may at first glance appear to resemble the frequency spectra of slowly pulsating B (SPB) stars (e.g. \citealt{Bowman2019, sharma22}), we present several reasons why this is not the case. SPB stars are mid-to-late B-type dwarf stars and hence hotter and less evolved than $\alpha$ Sex. Moreover, SPB stars are typically fast rotators \citep{Pedersen2021} and $\alpha$ Sex is not. Indeed, $\alpha$ Sex does not have the characteristic flat-bottomed Fe II lines that Vega has, which is a fast rotator seen pole-on \citep{Hill2010}. If, for example, $\alpha$ Sex were an SPB star and lies within an extension of the SPB stars towards cooler temperatures and lower gravities, it would need to be pulsating in gravity or perhaps Rossby modes (for which the restoring force is the Coriolis force) rather than pressure modes. However, the low $v\,\sin\,i$ combined with $i \simeq 90$ degrees given that the system is eclipsing based on the indicative UV photometry eclipse, makes $\alpha$ Sex unlikely to have gravito(inertial) modes. This is because the frequencies of slowly rotating SPB pulsators are not as high as 2.5 c/d as seen in the TESS light curve of $\alpha$ Sex (see e.g. \citealt{Pedersen2021}). For gravity-mode frequencies to be so high in such a star, the impact of the Coriolis force would also need to be large, thus $\alpha$ Sex would need to be rapidly rotating. Finally, gravity modes are unlikely to exist above the fundamental radial pressure mode frequency that we derive in this work. Hence we conclude low-radial order pressure and/or mixed modes are the most likely identification of the pulsations based on the fundamental parameters of $\alpha$ Sex derived in this work.

On the issue of whether pressure mode pulsations are expected in a star such as $\alpha$~Sex, we refer to \citet{Bowman2018} and \citet{Murphy2019}. In these studies of thousands of $\delta$~Sct stars observed by the Kepler space telescope, the observational hot edge of the classical instability strip was determined to correspond to $T_{\rm eff} \simeq 9000$~K based on the density of such stars in the Kiel and HR~diagrams. However, there exist a non-negligible number of outliers with hotter $T_{\rm eff}$ values. Pulsation excitation models struggle to explain the observed pulsations in such stars, because the heat-engine ($\kappa$) mechanism is inefficient at these temperatures \citep{Dupret2004}. 
Indeed, other pulsation excitation mechanisms operate in delta Scuti stars. \cite{antoci14} and \cite{antoci19} describe the role of turbulent pressure in the excitation of pulsations in $\delta$ Scuti stars, but such a mechanism typically excites high-radial order and thus high-frequency pressure modes. The discovery of pulsations in $\alpha$~Sex make it an interesting case study for follow-up asteroseismic modeling. Further monitoring with TESS may confirm the presence of eclipses too.

\acknowledgments 
{\it Acknowledgments.} The authors gratefully acknowledge Dr. Fr\'ed\'eric Arenou for enlightening discussion on Gaia parallaxes.
The authors thank the TESS science team for the excellent data, which were obtained from the Mikulski Archive for Space Telescopes (MAST) at the Space Telescope Science Institute (STScI), which is operated by the Association of Universities for Research in Astronomy, Inc., under NASA contract NAS5-26555. Support to MAST for these data is provided by the NASA Office of Space Science via grant NAG5-7584 and by other grants and contracts. Funding for the TESS mission is provided by the NASA Explorer Program.
This research has made use of the SIMBAD database, operated at CDS, Strasbourg, France.
This work was supported by FCT/MCTES through the research grants UIDB/04434/2020, UIDP/04434/2020 and PTDC/FIS-AST/30389/2017. DMB gratefully acknowledges a senior postdoctoral fellowship from the Research Foundation Flanders (FWO) with grant agreement no.~1286521N. MD is supported by national funds through FCT in the form of a work contract.

\appendix


\begin{center}
\begin{longtable}{cccccc}
\caption{Line identifications  for $\alpha$ Sex}\label{tab:ident}\\
\hline
 $\lambda_{\mathrm{obs}}$ (\AA)  & $\lambda_{\mathrm{lab}}$ (\AA) & Species & $\log gf$ & $E_{low}$ & Ref. \\
 \endfirsthead
 \multicolumn{6}{c}%
{\tablename\ \thetable\ -- \textit{Continued from previous page}} \\
\hline
$\lambda_{\mathrm{obs}}$ (\AA)  & $\lambda_{\mathrm{lab}}$ (\AA) & Species & $\log gf$ & $E_{low}$ & Ref. \\
\hline
\endhead
\hline \multicolumn{6}{c}{\textit{Continued on next page}} \\
\endfoot
\hline
\endlastfoot
  \hline
  
3900.44  & 3900.55 & \ion{Ti}{2} & -0.45      &  9118.260         &  VALD         \\
3903.78  & 3903.76 & \ion{Fe}{2} & -1.49      &  60807.23         &  VALD          \\
3905.97  & 3906.04 & \ion{Fe}{2} & -1.83      &  44929.549         &  VALD          \\
3913.47  & 3913.47 & \ion{Ti}{2} &  -0.53     & 8997.710          &  VALD           \\
3914.56  & 3914.50 & \ion{Fe}{2} & -4.05      &  13474.411      &   VALD           \\
3917.28  & 3917.32 & \ion{Mn}{2} & -1.15      &  55759.270         & VALD             \\
3918.32  & 3918.30 & \ion{Fe}{2} &  -3.63     &  73054.879         & VALD             \\
3918.76  & 3918.77 & \ion{Fe}{2} & -2.30      &  79246.171         &  VALD            \\	    
3920.59 & 3920.64 & \ion{Fe}{2} & -1.330 & 60628.698 & VALD \\ 
3923.43 & 3923.46 & \ion{S}{2} & 0.440 & 130641.115 & VALD \\ 
3924.953 & 3924.842 & \ion{Fe}{2} & -1.100 & 78138.990 & VALD \\ 
3926.527 & 3926.416 & \ion{Mn}{2} & -1.600 &  55759.270         &  VALD         \\  
3930.28 & 3930.30 & \ion{Fe}{1} & -1.490 & 704.007 & VALD \\ 
3933.71 & 3933.66  & \ion{Ca}{2} &  0.13     &  0.000       & VALD  \\
3936.07 & 3935.96 & \ion{Fe}{2} & -1.720 & 44915.056 & VALD \\ 
3938.29 & 3938.29 & \ion{Fe}{2} & -4.100 & 13474.447 & VALD \\ 
3939.02 & 3938.97 & \ion{Fe}{2} & -1.930 & 47674.729 & VALD \\ 
3941.18 & 3941.231 & \ion{Mn}{2} & -2.430 & 43852.362 & VALD \\ 
3943.84 & 3943.86 & \ion{Mn}{2} & -2.260 & 43699.122 & VALD \\ 
3947.33 & 3947.30 & \ion{O}{1} & -2.100 & 73768.202 & VALD \\ 
3990.95  & 3990.91  & \ion{S}{2}  & -0.50  & 128233.19  &   VALD \\ 
3993.52  & 3993.50  & \ion{S}{2}  & -0.82  & 115285.608  &VALD\\
3995.04  & 3995.00  & \ion{N}{2}  & 0.28   & 149187.803  &   VALD        \\
 4002.42 & 4002.54 & \ion{Fe}{2} & -2.070 & 48038.381 & VALD \\ 
4009.34  & 4009.26 & \ion{He}{1}  &  -1.47      &171134.998 &  VALD          \\
4012.53  & 4012.50  & \ion{Cr}{2} & -0.89       & 45669.369  & VALD           \\  
4024.67 & 4024.552 & \ion{Fe}{2} & -2.400 & 36252.930 &  VALD  \\ 
4026.26  & 4026.18 & \ion{He}{1}  & -2.62      & 169086.769 & hfs \\
         & 4026.19 & \ion{He}{1} & -0.70       & 169086.867 &   \\
	 & 4026.19 & \ion{He}{1} & -1.45       & 169086.769 &    \\
	 & 4026.20 & \ion{He}{1} & -0.98      & 169086.845  &   \\
	 & 4026.20 & \ion{He}{1} & -0.98      & 169086.845  &   \\
4032.94 & 4032.940 & \ion{Fe}{2} & -2.570 & 36254.622 & VALD \\ 
4045.85 & 4045.81 & \ion{Fe}{1} & 0.280 & 11976.239 & VALD \\
4048.86 & 4048.83 & \ion{Fe}{2} & -2.380 & 44917.017 & VALD \\ 
4057.43 & 4057.46 & \ion{Fe}{2} & -1.680 & 58668.776 & VALD \\
4061.80  & 4061.78 & \ion{Fe}{2}  & -2.650      & 48039.087          & VALD \\ 
4063.60 & 4063.60 & \ion{Fe}{1} & 0.060 & 12560.934 & VALD \\ 
4067.06 & 4067.03 & \ion{Ni}{2} & -1.830 & 32496.075 & VALD \\ 
4075.63  &  4075.62 & \ion{Cr}{2} & -3.470 &  25035.399 &  VALD \\ 
4081.39 & 4081.44 & \ion{Mn}{2} & -2.190 & 49288.543 & VALD \\ 
4111.85 & 4111.88 & \ion{Fe}{2} & -2.670 & 48038.381 & VALD \\ 
4119.62 & 4119.62  & \ion{Fe}{2} & -2.690 & 90780.019 & VALD  \\
4120.86  & 4120.81  & \ion{He}{1} & -1.740 & 169086.867 & hfs \\ 
         & 4120.82  & \ion{He}{1} & -1.960 & 169086.943 &  \\
	 & 4120.99  & \ion{He}{1} & -2.430 & 169087.931 &  \\ 
4122.63 & 4122.67 & \ion{Fe}{2} & -3.300 & 20830.553 & VALD \\ 
4124.67 & 4124.61 & \ion{Fe}{2} & -4.200 & 20516.953 & VALD \\ 
4128.06 & 4128.07 & \ion{Si}{2} & 0.360 & 79338.502 & VALD \\ 
4130.78 & 4130.872 & \ion{Si}{2} & -0.780 & 79355.019 & VALD \\ 
         & 4130.89 & \ion{Si}{2} & 0.550 & 79355.019 & VALD \\ 
4136.90 & 4136.902 & \ion{Mn}{2} & -1.250 & 49514.374 & VALD \\ 
4143.90 & 4143.76 & \ion{He}{1}&-1.200& 179134.998 &VALD\\ 
 4162.63 & 4162.67 & \ion{S}{2} & 0.780 & 128599.162 & VALD \\ 
4163.54 & 4163.65 & \ion{Ti}{2} & -0.130 & 20891.790 & VALD \\ 
4164.92 & 4164.92 & \ion{Fe}{3} &  1.010      & 198821.396 & VALD          \\
4167.21 & 4167.30 & \ion{Fe}{2} &  -0.560     &  90300.626  &  VALD          \\
4169.04 & 4168.97 & \ion{He}{1} &  -2.340     &  171134.998 & VALD             \\    
4171.02 & 4171.03 & \ion{Mn}{2} & -2.340 & 49425.654 & VALD \\ 
4171.85 & 4171.903 & \ion{Cr}{2} & -2.940 & 25043.517 & VALD \\ 
        & 4171.904 & \ion{Ti}{2} & -0.290 & 20951.754 & VALD \\ 
4173.46 & 4173.46 & \ion{Fe}{2} & -2.160 & 20830.553 & VALD \\ 
4174.17 & 4174.27 & \ion{S}{2} & 0.800 & 140319.232 & VALD \\ 
4175.97  & 4175.99  & \ion{Fe}{2} &  -3.840     & 54902.315  & VALD \\ 
4178.61 & 4178.63 & \ion{Fe}{2} & -4.290  & 60402.341 & VALD\\ 
4199.39 & 4199.49 & \ion{Fe}{2} & -0.330 & 89922.758 & VALD \\ 
4200.47 & 4200.518 & \ion{Fe}{2} & -0.410 & 90067.932 & VALD \\ 
4205.38 & 4205.38 & \ion{Mn}{2} & -3.450 & 14593.834 & VALD \\ 
4206.44 & 4206.37 & \ion{Mn}{2} & -1.540 & 43528.661 & VALD \\ 
4233.12 & 4233.17 & \ion{Fe}{2} & -1.810 & 20830.553 & VALD \\ 
4239.05 & 4239.15 & \ion{Mn}{2} & -2.240 & 43311.972 & VALD \\ 
4242.33 & 4242.33 & \ion{Mn}{2} & -1.260 & 49820.865 & VALD \\ 
         & 4242.36 & \ion{Cr}{2} & -1.170 & 31221.723 & VALD \\
4244.46 & ?  & &  &  &  \\ 
4246.85 & 4246.82 & \ion{Sc}{2} & 0.240 & 2540.950 & VALD \\ 
4250.43 & 4250.44 & \ion{Fe}{2} & -1.720 & 61974.931 & VALD \\ 
4251.64 & 4251.72 & \ion{Mn}{2} & -1.060 & 49885.389 & VALD \\ 
4252.97 & 4252.96 & \ion{Mn}{2} & -1.140 & 49893.458 & VALD \\ 
4258.17 & 4258.11 & \ion{Fe}{2} & -3.500 & 21812.044 & VALD \\ 
4259.17 & 4259.17 & \ion{Mn}{2} & -1.440 & 43537.186 & VALD \\ 
4261.92 & 4261.91 & \ion{Cr}{2} & -1.340 & 31165.263 & VALD \\ 
4263.81 & 4263.87 & \ion{Fe}{2} & -1.690 & 62048.230 & VALD \\ 
4267.15 & 4267.18 & \ion{C}{2} & 0.720 & 145550.705 & VALD \\ 
4273.26 & 4273.33 & \ion{Fe}{2} & -3.300 & 21812.044 & VALD \\ 
4278.29 & 4278.15 & \ion{Fe}{2} & -3.950 & 21712.445 & VALD \\ 
4286.17  & 4286.16  & \ion{Fe}{3} & 0.700 & 184316.573 & VALD   \\      
4288.05  & 4288.07  & \ion{Mn}{2} & -2.760 &  43311.301 & VALD          \\
4290.25 & 4290.22 & \ion{Ti}{2} & -0.850 & 9395.802 & VALD \\ 
4292.17 & 4292.23 & \ion{Mn}{2} & -1.540 & 43395.394 & VALD \\ 
4294.18 & 4294.10 & \ion{Ti}{2} & -1.110 & 8744.341 & VALD \\ 
4296.57 & 4296.57 & \ion{Fe}{2} & -2.900 & 21812.044 & VALD \\ 
4300.18 & 4300.04 & \ion{Ti}{2} & -0.440 & 9518.152 & VALD \\ 
4303.10 & 4303.18 & \ion{Fe}{2} & -2.610 & 21812.044 & VALD \\ 
4307.92 & 4307.87 & \ion{Ti}{2} & -1.070 & 9395.802 & VALD \\ 
4312.98 & 4312.85 & \ion{Ti}{2} & -1.100 & 9518.152 & VALD \\ 
4314.24 & 4314.31 & \ion{Fe}{2} & -3.600 & 21583.397 & VALD \\ 
4320.92 & 4320.96 & \ion{Ti}{2} & -1.800 & 9395.802 & VALD \\ 
4325.44 & 4325.44 & \ion{Fe}{2} & -2.380 & 49506.310 & VALD \\ 
        & 4325.53 & \ion{Fe}{2} & -2.360 & 49103.030 & VALD \\ 
4326.59 & 4326.64 & \ion{Mn}{2} & -1.370 & 43537.186 & VALD \\ 
4351.70 & 4351.76 & \ion{Fe}{2} & -2.080 & 21812.044 & VALD \\ 
4354.39 & 4354.34 & \ion{Fe}{2} & -1.350 & 61725.611 & VALD \\ 
4357.61 & 4357.58 & \ion{Fe}{2} & -2.010 & 49100.956 & VALD \\ 
4361.15 & 4361.25 & \ion{Fe}{2} & -2.260 & 49506.310 & VALD \\ 
4363.16 & 4363.26 & \ion{Mn}{2} & -1.890 & 44900.887 & VALD \\ 
4365.20 & 4365.22 & \ion{Mn}{2} & -1.340 & 53014.822 & VALD \\ 
4368.21 & 4368.242 & \ion{O}{1} & -1.960 & 76794.977 & VALD \\ 
         & 4368.250 & \ion{O}{1} & -2.190 & 76794.977 & VALD \\ 
4369.42 & 4369.400 & \ion{Fe}{2} & -3.600 & 22409.818 & VALD \\ 
4374.71  & 4374.82  & \ion{Ti}{2} & -1.290 & 16625.110 &  VALD         \\   
4377.61 &  & ? &   &  &  \\ 
4379.62 & 4379.67 & \ion{Mn}{2} & -1.870 & 43852.362 & VALD \\ 
4382.51 & 4382.51  & \ion{Fe}{3} & -3.020 &  66522.952 &  VALD         \\
4384.40 & 4384.32 & \ion{Fe}{2} & -3.700 & 21430.357 & VALD \\ 
4385.35 & 4385.39 & \ion{Fe}{2} & -2.600 & 22409.818 & VALD \\ 
 4388.04 & 4387.93 & \ion{He}{1} & -0.887 & 171134.897 & VALD \\ 
4390.55 & 4390.514 & \ion{Mg}{2} & -1.480 & 80650.022 & VALD \\ 
         & 4390.564 & \ion{Mg}{2} & -0.520 & 80650.022 & VALD \\ 
4395.04 & 4395.031 & \ion{Ti}{2} & -0.660 & 8744.341 & VALD \\ 
4395.83  & 4395.76  & \ion{Fe}{3} & -2.600 & 66591.679 &   VALD       \\
4402.92 & 4402.879 & \ion{Fe}{2} & -2.600 & 49506.924 & VALD \\ 
4404.87 & 4404.75 & \ion{Fe}{1} & -0.140 & 12560.934 & VALD \\ 
4407.74 & 4407.68 & \ion{Fe}{2} & 0.780 & 109473.634 & VALD \\ 
4409.54 & 4409.53 & \ion{Fe}{2} & 0.950 & 109489.771 & VALD \\ 
4411.14 & 4411.072 & \ion{Ti}{2} & -0.670 & 24961.191 & VALD \\ 
        & 4411.15   & \ion{C}{2} & 0.530 & 198425.425 &   VALD \\
4413.62 & 4413.60 & \ion{Fe}{2} & -4.200 & 21581.615 & VALD \\ 
4416.81 & 4416.83 & \ion{Fe}{2} & -2.600 & 22409.818 & VALD \\ 
4417.70 & 4417.71 & \ion{Ti}{2} & -1.430 & 9395.802 & VALD \\ 
4418.73 & 4418.78 & \ion{Ce}{2} & 0.180       & 6967.547 &   VALD        \\
4419.56 & 4419.60 & \ion{Fe}{3} & -2.210 & 66468.153 & VALD \\
        & 4419.60 & \ion{Cr}{2} & -0.260 & 94365.189 &   VALD         \\
4428.01 & 4427.99 & \ion{Mg}{2} & -1.210 & 80619.500 & VALD \\ 
4430.99 & 4431.02 & \ion{Fe}{3} &  -2.570 & 66522.952          &  VALD         \\ 
4434.00 & 4433.99 & \ion{Mg}{2} & -0.910 & 80650.022 & VALD \\ 
4437.58  & 4437.55 & \ion{He}{1} & -2.030 & 171134.998 & VALD  \\
4443.81 & 4443.79 & \ion{Ti}{2} & -0.720 & 8710.567 & VALD \\ 
4451.49 & 4451.55 & \ion{Fe}{2} & -1.910 & 49506.924 & VALD \\ 
4455.33 & 4455.27 & \ion{Fe}{2} & -2.000 & 50216.076 & VALD \\ 
4461.65 & 4461.71 & \ion{Fe}{2} & -2.100 & 50212.834 & VALD \\ 
4463.43  & 4463.41  & \ion{Ce}{2} & -0.110 & 7722.285& VALD           \\
4464.46  & 4464.45  & \ion{Ti}{2} & -2.080 &  9363.620 &  VALD           \\
4468.51 & 4468.51 & \ion{Ti}{2} & -0.620 & 9118.285 & VALD \\ 
4469.93  & 4469.90  & \ion{Mn}{2} & -3.170     & 49882.153  &          \\
4471.46 & 4471.470 & \ion{He}{1} & -2.210  & 169086.769 & hfs \\
        & 4471.474 & \ion{He}{1} & -1.040  & 169086.769 &            \\
	& 4471.474 & \ion{He}{1} & -0.240  & 169086.769 &            \\
        & 4471.486 & \ion{He}{1} & -1.040  & 169086.769 &            \\
	& 4471.489 & \ion{He}{1} & -0.560  & 169086.769 &            \\ 
4472.73 & 4472.62 & \ion{Fe}{2} & -2.340 & 61974.931 &     \\ 
4478.67 & 4478.64 & \ion{Mn}{2} & -0.940 & 53595.541 & VALD \\ 
4481.21 & 4481.130 & \ion{Mg}{2} & 0.750 & 71490.190 & VALD \\ 
        & 4481.150 & \ion{Mg}{2} & -0.550 & 71490.190 & VALD \\ 
        & 4481.327 & \ion{Mg}{2} & 0.590 & 71491.058 & VALD \\ 
4483.49 & 4483.43  & \ion{S}{2}  & -0.320 & 128233.197  &  VALD         \\	
4489.19 & 4489.19 & \ion{Fe}{2} & -3.000 & 22810.345 & VALD \\ 
4491.37 & 4491.40 & \ion{Fe}{2} & -2.640 & 23031.283 & VALD \\ 
4493.47 & 4493.53 & \ion{Fe}{2} & -1.560 & 63876.325 & VALD \\ 
4494.41 & 4494.42 & \ion{Zr}{2} & -0.480 & 19433.239 & VALD \\ 
4501.24 & 4501.27 & \ion{Ti}{2} & -0.770 & 8997.787 & VALD \\ 
        & 4501.28  & \ion{Fe}{2} &-0.950 & 90387.870 &  VALD          \\ 
4515.39 & 4515.34 & \ion{Fe}{2} & -2.360 & 22939.352 & VALD \\ 
4518.91 & 4518.96 & \ion{Mn}{2} & -1.310 & 53595.541 & VALD \\ 
4520.15 & 4520.22 & \ion{Fe}{2} & -2.600 & 22637.195 & VALD \\ 
4522.57 & 4522.63 & \ion{Fe}{2} & -1.990 & 22939.352 & VALD \\ 
4524.87 & 4524.94 & \ion{S}{2} & 0.030 & 121530.021 & VALD \\ 
4526.50  & 4526.40  & \ion{Fe}{2} & -2.330       & 62962.204  &  VALD  \\
4529.28  &   ?      &             &        &          &           \\
4534.00 & 4533.97 & \ion{Ti}{2} & -0.770 & 9975.999 & VALD \\
4541.45 & 4541.52 & \ion{Fe}{2} & -3.000 & 23031.283 & VALD \\ 
4549.39 & 4549.47 & \ion{Fe}{2} & -1.730 & 22810.345 & VALD \\ 
4552.47 & 4552.41 & \ion{S}{2}  & -0.100 & 121528.718  & VALD         \\
4555.01 & 4554.99 & \ion{Cr}{2} & -1.370 & 32836.653 & VALD \\ 
4555.86 & 4555.89 & \ion{Fe}{2} & -2.250 & 22810.345 & VALD \\ 
4558.61 & 4558.64 & \ion{Cr}{2} & -0.660 & 32854.249 & VALD \\ 
4563.69 & 4563.76 & \ion{Ti}{2} & -0.960 & 9851.014 & VALD \\ 
4571.96 & 4571.97 & \ion{Ti}{2} & -0.320 & 12677.106 & VALD \\ 
4576.34 & 4576.33 & \ion{Fe}{2} & -2.920 & 22939.352 & VALD \\
4579.67  & 4579.72  & \ion{Ti}{2} & -1.950 & 40581.490 &  VALD         \\
4582.69 & 4582.84 & \ion{Fe}{2} & -3.060 & 22939.352 & VALD \\ 
4583.81 & 4583.84 & \ion{Fe}{2} & -1.740 & 22637.195 & VALD \\ 
4588.12  & 4588.20 & \ion{Cr}{2} & -0.640 & 32836.653 & VALD \\ 
4589.87 & 4589.95 & \ion{Ti}{2} & -1.780 & 9975.999 & VALD \\ 
4595.95 & 4596.02 & \ion{Fe}{2} & -1.960 & 50212.834 & VALD \\ 
4598.43 & 4598.49 & \ion{Fe}{2} & -1.540 & 62945.045 & VALD \\ 
4616.63 & 4616.63 & \ion{Cr}{2} & -1.290 & 32844.706 & VALD \\ 
4618.79 & 4618.81 & \ion{Cr}{2} & -1.110 & 32854.941 & VALD \\ 
4620.50 & 4620.51 & \ion{Fe}{2} & -3.190 & 22810.345 & VALD \\ 
4621.51 & 4621.40 & \ion{N}{2}  & -0.540  & 148940.167 & VALD \\ 
4625.70  & 4625.64  & \ion{C}{2}  & 0.700  & 199941.414  &  VALD          \\
4629.31 & 4629.29 & \ion{Ti}{2} & -2.250 & 9518.152 & VALD \\ 
        & 4629.332 & \ion{Fe}{2} & -2.260 & 22637.195 & VALD \\ 
4634.09 & 4634.07 & \ion{Cr}{2} & -1.240 & 32844.706 & VALD \\ 
4635.21 & 4635.32 & \ion{Fe}{2} & -1.580 & 48039.109 & VALD \\ 
4638.05 & 4638.04 & \ion{Fe}{2} & -1.540 & 62169.219 & VALD \\ 
4656.90 & 4656.98 & \ion{Fe}{2} & -3.570 & 23317.635 & VALD \\ 
4663.34  &  ?       &             &                   &           \\
4713.31  & 4713.38  & \ion{He}{1} & -1.920 & 169087.931  &  VALD  \\
4716.40 & 4716.267 & \ion{S}{2} & -0.370 & 109831.595 & VALD \\ 
4727.92 & 4727.84 & \ion{Mn}{2} & -1.950 & 43311.972 & VALD \\ 
4730.37 & 4730.40 & \ion{Mn}{2} & -2.020 & 43336.173 & VALD \\ 
4731.49 & 4731.44 & \ion{Fe}{2} & -3.100 & 23317.635 & VALD \\ 
4738.23 & 4738.29 & \ion{Mn}{2} & -1.870 & 43395.394 & VALD \\ 
4755.72 & 4755.73 & \ion{Mn}{2} & -1.220 & 43529.741 & VALD \\ 
4764.64 & 4764.728 & \ion{Mn}{2} & -1.330 & 43537.810 & VALD \\ 
4784.52 & 4784.63& \ion{Mn}{2} & -1.500 & 53014.822 & VALD \\ 
4791.82 & 4791.78 & \ion{Mn}{2} & -1.680 & 49885.389 & VALD \\ 
4806.71 & 4806.860 & \ion{Mn}{2} & -1.840 & 43696.217 & VALD \\ 
4815.51 & 4815.55 & \ion{S}{2} & 0.070 & 110268.595 & VALD \\ 
        & 4815.49 & \ion{Mn}{2}  & -3.780    & 43696.190 &          \\
4824.11  & 4824.06  & \ion{S}{2}  & 0.180  & 110268.595  &  VALD        \\
         & 4824.13  & \ion{Cr}{2} & -1.230 & 31219.331 & VALD \\ 
4826.53  & 4826.51 & \ion{Mn}{2} & -3.000  & 94230.896 &  \\ 
4848.18 & 4848.24 & \ion{Cr}{2} & -1.130 & 31168.575 & VALD \\ 
4876.37 & 4876.40 & \ion{Cr}{2} & -1.460 & 31084.610 & VALD \\ 
4883.28 & 4883.28 & \ion{Fe}{2} & -0.600 & 82853.704 & VALD \\ 
4908.12 & 4908.15 & \ion{Fe}{2} & -0.270 & 83308.242 & VALD \\
4909.75  & 4909.75  & \ion{Fe}{2} & -2.610 & 82978.679 & VALD           \\ 
4911.52  &    ?     &             &        &           &           \\   
4913.29 & 4913.30 & \ion{Fe}{2} & 0.050 & 82978.717 & VALD \\ 
4917.22 & 4917.21 & \ion{S}{2} & -0.380 & 112937.572 & VALD \\ 
4921.86  &    ?     &             &        &           &          \\      
4923.90 & 4923.92 & \ion{Fe}{2} & -1.210 & 23317.635 & VALD \\ 
4948.42 & 4948.49  & \ion{Mn}{2}  & -3.220      &  62572.200  &   VALD  \\ 
4951.58 & 4951.58 & \ion{Fe}{2} & 0.210 & 83136.510 & VALD \\ 
4954.03 & 4953.98 & \ion{Fe}{2} & -2.810 & 44933.149 & VALD \\ 
4977.02 & 4977.04 & \ion{Fe}{2} & -0.040 & 83558.566 & VALD \\ 
4984.52 & 4984.49 & \ion{Fe}{2} & 0.080 & 83308.242 & VALD \\ 
        & 4984.58 & \ion{Fe}{2} & -1.260 & 83990.065 & VALD           \\
4990.51 & 4990.51 & \ion{Fe}{2} & 0.200 & 83308.242 & VALD \\ 
4991.70  &  ?     &             &       &           &            \\ 
4993.38 & 4993.36 & \ion{Fe}{2} & -3.700 & 22637.195 & VALD \\ 
5001.86 & 5001.95 & \ion{Fe}{2} & 0.920 & 82853.704 & VALD \\ 
5004.19 & 5004.19 & \ion{Fe}{2} & 0.500 & 82853.704 & VALD \\ 
5007.21  &    ?     &             &        &           &          \\ 
5009.54 & 5009.56 & \ion{S}{2} & -0.230 & 109831.595 & VALD \\ 
5014.03 & 5014.04 & \ion{S}{2} & 0.050 & 113461.537 & VALD \\ 
5015.63 & 5015.68 & \ion{He}{1} & -0.820 & 166277.440 & VALD \\ 
        & 5015.75 & \ion{Fe}{2} & -0.030 & 83459.720 & VALD \\ 
5018.37 & 5018.44 & \ion{Fe}{2} & -1.350 & 23317.635 & VALD \\ 
5019.70  &   ?    &             &        &           &           \\
5021.42  & 5021.47  & \ion{Ni}{2} & 0.920 & 100309.295 &  VALD         \\
5022.60 & 5022.42 & \ion{Fe}{2} & -0.070 & 83459.720 & VALD \\ 
5026.91 & 5026.80 & \ion{Fe}{2} & -0.440 & 83136.510 & VALD \\ 
5028.95  & 5028.98 & \ion{Ni}{2}  & 0.170 &  100389.516        & VALD  \\
5030.63 & 5030.63 & \ion{Fe}{2} & 0.430 & 82978.717 & VALD \\
5032.58 & 5032.45 & \ion{S}{2} & 0.190 & 110268.595 & VALD \\ 
5034.21 & 5034.17 & \ion{Fe}{2} & -0.800 & 83558.566 & VALD \\ 
5035.69 & 5035.71 & \ion{Fe}{2} & 0.630 & 82978.717 & VALD \\ 
5036.856 & 5036.713 & \ion{Fe}{2} & -0.560 & 83812.350 & VALD \\ 
5040.92 & 5041.02 & \ion{Si}{2} & 0.030 & 81191.341 & VALD \\ 
5045.11 & 5045.11 & \ion{Fe}{2} & 0.000 & 83136.510 & VALD \\ 
5047.62 & 5047.64 & \ion{Fe}{2} & -0.240 & 83136.510 & VALD \\
        & 5047.74 & \ion{He}{1} & -1.600 & 171134.198  &  VALD          \\  
5056.07 & 5055.98 & \ion{Si}{2} & 0.520 & 81251.320 & VALD \\ 
5061.74 & 5061.72 & \ion{Fe}{2} & 0.280 & 83136.510 & VALD \\ 
5067.89 & 5067.890 & \ion{Fe}{2} & -0.080 & 83308.242 & VALD \\ 
5070.90 & 5070.90 & \ion{Fe}{2} & 0.270 & 83136.510 & VALD \\ 
5073.94 & 5073.896 & \ion{Fe}{2} & -0.720 & 84266.595 & VALD \\ 
        & 5074.051 & \ion{Fe}{2} & -2.200 & 54904.243 & VALD \\ 
	& 5073.90  & \ion{Fe}{3} & -2.560 &  69788.188 & VALD \\
5075.80 & 5075.760 & \ion{Fe}{2} & 0.180 & 84326.965 & VALD \\ 
5082.10 & 5082.23 & \ion{Fe}{2} & -0.130 & 83990.103 & VALD \\ 
5087.33 & 5087.300 & \ion{Fe}{2} & -0.420 & 83713.585 & VALD \\ 
        & 5087.419 & \ion{Y}{2} & -0.170 & 8743.050 & VALD \\ 
5089.25 & 5089.21 & \ion{Fe}{2} & 0.010 & 83308.242 & VALD \\ 
5093.62 & 5093.58 & \ion{Fe}{2} & -2.320 & 54869.899 & VALD \\ 
5097.19 & 5097.27 & \ion{Fe}{2} & 0.320 & 83713.585 & VALD \\ 
5098.70 & 5098.68 & \ion{Fe}{2} & -0.490 & 84268.805 & VALD \\ 
5100.71 & 5100.73 & \ion{Fe}{2} & 0.720 & 83726.416 & VALD \\ 
5102.54 & 5102.52 & \ion{Mn}{2} & -1.900 & 48320.677 & VALD \\ 
5106.12 & 5106.11 & \ion{Fe}{2} & -0.250 & 83308.242 & VALD \\ 
5107.33  & 5107.25  & \ion{Fe}{2} & -3.960 &  50075.908 & VALD         \\
5116.99 & 5117.031 & \ion{Fe}{2} & -0.040 & 84131.623 & VALD \\ 
5123.31 & 5123.33 & \ion{Mn}{2} & -1.850 & 48336.810 & VALD \\ 
5127.53  & 5127.63  & \ion{Fe}{3} & -2.560  & 69837.758  &  VALD         \\
5129.96  &  ?       &             &       &            &           \\
5132.58 & 5132.66 & \ion{Fe}{2} & -4.100 & 22637.195 & VALD \\ 
5133.97 & 5134.07 & \ion{Fe}{2} & 0.290 & 104916.599 & VALD \\ 
5144.28 & 5144.32 & \ion{Fe}{2} & 0.310 & 84424.421 & VALD \\ 
5145.85 & 5145.730 & \ion{Fe}{2} & -0.210 & 83990.103 & VALD \\ 
        & 5145.822 & \ion{Fe}{2} & -0.140 & 83990.103 & VALD \\ 
5149.25 & 5149.19 & \ion{Fe}{2} & 0.410 & 103601.919 & \\ 
5152.850 & 5152.88 & \ion{Fe}{2} & -0.990 & 83990.065 &VALD\\ 
5154.24  &    ?     &            &    &   &   \\
5156.31  & 5156.28  & \ion{Fe}{2} & -3.750   & 93487.649 & \\ 
5157.40 & 5157.28 & \ion{Fe}{2} & -0.170 & 84326.965 & VALD \\ 
5160.83 & 5160.840 & \ion{Fe}{2} & -2.600 & 44915.056 & VALD \\ 
5163.37  &   ?      &             &       &            &           \\  
5166.17 & 5166.20 & \ion{Fe}{2} & 0.930 & 108134.747 & VALD \\ 
5168.92 & 5169.03 & \ion{Fe}{2} & -0.870 & 23317.635 & VALD \\ 
5170.49  &   ?    &             &        &           &           \\   
5172.97 & 5173.13 & \ion{Fe}{2} & 0.460 & 105061.784 & VALD \\ 
5177.37 & 5177.37 & \ion{Fe}{2} & 1.160 & 103642.253 & VALD \\ 
5180.21 & 5180.31 & \ion{Fe}{2} & -0.090 & 83812.350 & VALD \\ 
5183.57 & 5183.53 & \ion{Fe}{2} & -0.850 &  84131.564 &VALD\\  
5188.80 & 5188.687 & \ion{Ti}{2} & -1.220 & 12758.260 & VALD \\ 
5189.96 & 5190.00 & \ion{Fe}{2} & 0.490 & 105400.535 & VALD \\ 
 5192.70 & 5192.62 & \ion{Fe}{2} & 0.200 & 103101.857 & VALD \\ 
5196.01 & 5195.94 & \ion{Fe}{2} & 0.920 & 104174.580 & VALD \\ 
5197.55 & 5197.57 & \ion{Fe}{2} & -2.050 & 26055.412 & VALD \\ 
5199.00 & 5199.12 & \ion{Fe}{2} & 0.120 & 83713.585 & VALD \\ 
5200.92 & 5200.80 & \ion{Fe}{2} & -0.040 & 83812.350 & VALD \\ 
        & 5201.03 & \ion{S}{2}  & 0.430  & 121528.718 & VALD          \\
5210.58 & 5210.550 & \ion{Fe}{2} & 0.790 & 103166.386 & VALD \\ 
5212.61 & 5212.62 & \ion{S}{2} & 0.320 & 121530.021 & VALD \\ 
5213.97 & 5213.95 & \ion{Fe}{2} & -0.260 & 84527.806 & VALD \\ 
        & 5214.003 & \ion{Fe}{2} & 0.860 & 104069.717 & VALD \\ 
5215.48 & 5215.35 & \ion{Fe}{2} & 0.000 & 83713.585 & VALD \\ 
5216.90 & 5216.86 & \ion{Fe}{2} & 0.670 & 84527.806 & VALD \\ 
        & 5216.859 & \ion{Fe}{2} & 0.480 & 84710.751 & VALD \\ 
5218.70 & 5218.84 & \ion{Fe}{2} & -0.170 & 83726.416 & VALD \\ 
5223.26 & 5223.20 & \ion{Fe}{2} & 0.450 & 103876.141 & VALD \\ 
        & 5223.260 & \ion{Fe}{2} & -0.170 & 83812.350 & VALD \\ 
5225.26 & 5225.21 & \ion{Fe}{2} & 0.980 & 105287.615 & VALD \\ 
         & 5225.35 & \ion{Fe}{2} & 0.710 & 103190.571 & VALD \\ 
5227.44 & 5227.32 & \ion{Fe}{2} & 0.190 & 84844.864 & VALD \\ 
        & 5227.49 & \ion{Fe}{2} & 0.850 & 84296.883 & VALD \\ 
5228.71 & 5228.63 & \ion{Fe}{2} & 0.900 & 108779.995 & VALD \\ 
5232.76 & 5232.78 & \ion{Fe}{2} & -0.080 & 83726.416 & VALD \\ 
5234.53 & 5234.44 & \ion{Fe}{2} & -2.210 & 25981.646 & VALD \\ 
        & 5236.621 & \ion{Fe}{2} & -0.680 & 84326.965 & VALD \\ 
5237.90 & 5237.95 & \ion{Fe}{2} & 0.100 & 84266.595 & VALD \\ 
5239.62 & 5239.65  & \ion{Mn}{2} & -3.430 & 82419.479 &  \\   
5241.00 & 5241.06 & \ion{Fe}{2} & -0.580 & 83812.350 & VALD \\ 
5245.11 & 5245.07 & \ion{Fe}{2} & 0.870 & 108465.440 & VALD \\ 
5247.95 & 5247.96 & \ion{Fe}{2} & 0.550 & 84938.264 & VALD \\ 
5251.20 & 5251.21 & \ion{Fe}{2} & -0.660 & 84325.265 & VALD \\ 
        & 5251.23 & \ion{Fe}{2} & 0.420 & 84844.864 & VALD \\ 
5254.77 & 5254.920 & \ion{Fe}{2} & -3.340 & 26051.708 & VALD \\ 
5257.02 & 5256.93 & \ion{Fe}{2} & -4.180 & 23317.488 & VALD \\ 
        & 5257.12 & \ion{Fe}{2} & 0.160 & 84685.247 & VALD \\ 
5258.08 & 5258.03 & \ion{Fe}{2} & -2.100    & 84685.198 & VALD\\ 
        & 5258.12 & \ion{Fe}{2} &  -0.920    & 84424.372  & VALD \\
5260.15 & 5260.254 & \ion{Fe}{2} & 1.090 & 84035.172 & VALD \\ 
5262.462 & 5262.313 & \ion{Fe}{2} & -0.370 & 85048.655 & VALD \\ 
5264.169 & 5264.020 & \ion{Fe}{2} & -0.440 & 84685.247 & VALD \\ 
5264.328 & 5264.179 & \ion{Fe}{2} & 0.300 & 84710.751 & VALD \\ 
5264.364 & 5264.215 & \ion{Mg}{2} & -0.370 & 93310.590 & VALD \\ 
5264.49 & 5264.36 & \ion{Mg}{2} & -0.530 & 93311.112 & VALD \\ 
5272.38 & 5272.41 & \ion{Fe}{2} & -2.010 & 48039.109 & VALD \\ 
5274.83 & 5274.965 & \ion{Cr}{2} & -1.560 & 32834.831 & VALD \\ 
5275.92 & 5275.997 & \ion{Fe}{2} & -1.900 & 25805.326 & VALD \\ 
5284.02 & 5284.07 & \ion{Fe}{2} & -3.200 & 23317.635 & VALD \\ 
        & 5284.11  & \ion{Fe}{2} &  -3.190 & 23317.632  & VALD          \\ 
5291.67 & 5291.67  & \ion{Fe}{2} & 0.580 & 84527.779  & VALD    \\
5296.99 & 5297.000 & \ion{Mn}{2} & -0.240 & 79550.458 & hfs \\ 
       & 5297.028 & \ion{Mn}{2} & 0.400 & 79550.458 & \\ 
        & 5297.060 & \ion{Mn}{2} & 0.620 & 79550.502 & \\ 
5299.29 & 5299.302 & \ion{Mn}{2} & -0.440 & 79558.533 & hfs \\ 
        & 5299.330 & \ion{Mn}{2} & 0.380 & 79558.533 &  \\ 
        & 5299.370 & \ion{Mn}{2} & 0.830 & 79558.560 & \\ 
5302.39 & 5302.402 & \ion{Mn}{2} & 0.200 & 79566.596 & hfs \\ 
        & 5302.440 & \ion{Mn}{2} & 1.000 & 79569.219 &  \\ 
5306.06 & 5306.18 & \ion{Fe}{2} & 0.040 & 84870.912 & VALD \\ 
5308.36 & 5308.41 & \ion{Fe}{2} & 0.550 & 104118.119 & VALD \\ 
        & 5308.41 & \ion{Cr}{2} & -1.810 & 32836.653 & VALD \\ 
5316.62 & 5316.61 & \ion{Fe}{2} & -1.780 & 25428.789 & VALD \\ 
5316.78 & 5316.78 & \ion{Fe}{2} & -2.800 & 25981.646 & VALD \\ 
5318.21 & 5318.06 & \ion{Fe}{2} & -0.230 & 84527.806 & VALD \\ 
5320.68 & 5320.72 & \ion{S}{2} & 0.430 & 121530.021 & VALD \\ 
5322.01 & 5321.86 & \ion{Fe}{2} & 0.670 & 106021.587 & VALD \\ 
5325.38 & 5325.543 & \ion{Fe}{2} & -3.260 & 25981.646 & VALD \\ 
5330.78 & 5330.73 &  \ion{O}{1}  & -2.570 & 86631.453 &   VALD \\
        & 5330.74 & \ion{O}{1} & -1.570 & 86631.453 & VALD \\ 
        & 5330.740 & \ion{O}{1} & -0.980 & 86631.453 & VALD \\ 
	& 5330.78  & \ion{Ne}{1} & -1.040 & 148257.789 &    \\
5339.50 & 5339.594 & \ion{Fe}{2} & 0.520 & 84296.883 & VALD \\ 
 5362.75 & 5362.74 & \ion{Fe}{2} & -0.190 & 84710.751 & VALD \\ 
5370.43 & 5370.28 & \ion{Fe}{2} & -0.570 & 84266.595 & VALD \\ 
5375.83 & 5375.842 & \ion{Fe}{2} & -0.330 & 84296.883 & VALD \\ 
5375.994 & 5375.842 & \ion{Fe}{2} & -0.700 & 85048.655 & VALD \\ 
5387.00 & 5387.06 & \ion{Fe}{2} & 0.500 & 84863.380 & VALD \\ 
5393.80 & 5393.84 & \ion{Fe}{2} & -0.250 & 84296.883 & VALD \\ 
5395.83 & 5395.86 & \ion{Fe}{2} & 0.280 & 85495.368 & VALD \\ 
5401.95 & 5401.89 & \ion{Fe}{2} & -0.840 & 85462.859 &  \\ 
5405.14 & 5405.09 & \ion{Fe}{2} & -0.430 & 85172.826 & VALD \\ 
 5411.49 & 5411.373 & \ion{Fe}{2} & -0.050 & 85495.368 & VALD \\ 
5414.42  &   ?      &             &       &           &           \\      
5427.89 & 5427.83 & \ion{Fe}{2} & -1.580 & 54232.199 & VALD \\ 
5429.93 & 5429.987 & \ion{Fe}{2} & 0.430 & 85462.908 & VALD \\ 
5432.85 & 5432.82 & \ion{S}{2} & 0.200 & 109831.595 & VALD \\ 
5442.36 & 5442.359 & \ion{Fe}{2} & -0.310 & 85048.655 & VALD \\ 
5443.80 &    ?    &             &        &           &           \\  
5444.25 & 5444.39 & \ion{Fe}{2} & -0.170 & 85495.368 & VALD \\ 
5445.67 & 5445.80 & \ion{Fe}{2} & -0.110 & 85048.655 & VALD \\ 
5455.98 & 5455.930 & \ion{Fe}{2} & -0.510 & 84527.806 & VALD \\ 
5457.70 & 5457.72 & \ion{Fe}{2} & -0.140 & 85728.844 & VALD \\ 
5465.91 & 5465.932 & \ion{Fe}{2} & 0.350 & 85679.757 & VALD \\ 
5467.049 & 5466.894 & \ion{Si}{2} & -0.080 & 101024.349 & VALD \\ 
         & 5466.912 & \ion{Fe}{2} & -1.870 & 54902.288 & VALD \\ 
5473.15  & 5473.15  & \ion{Fe}{2} & -1.440 & 86844.832 &   VALD        \\
5475.89 & 5475.83 & \ion{Fe}{2} & -0.080 & 84685.247 & VALD \\ 
5478.32 & 5478.365 & \ion{Cr}{2} & -1.970 & 33697.843 & VALD \\ 
        & 5479.399 & \ion{Fe}{2} & -0.350 & 85172.826 & VALD \\ 
5479.44  & 5479.40 & \ion{Fe}{2} & -0.410       &  87172.810 & VALD  \\
5482.22 & 5482.31 & \ion{Fe}{2} & 0.410 & 85184.772 & VALD \\ 
5487.53 & 5487.62 & \ion{Fe}{2} & 0.290 & 85462.908 & VALD \\ 
5492.23 & 5492.078 & \ion{Fe}{2} & 0.090 & 85679.757 & VALD \\ 
5493.91 & 5493.83 & \ion{Fe}{2} & 0.260 & 84685.247 & VALD \\ 
5502.95  &  ?     &            &        &           &            \\
5506.20 & 5506.199 & \ion{Fe}{2} & 0.860 & 84863.380 & VALD \\ 
5510.87 & 5510.78 & \ion{Fe}{2} & 0.100 & 85184.772 & VALD \\
5529.47  & 5529.47  & \ion{Cr}{2} & -2.890 & 97899.412 & \\
5532.05 & 5532.09 & \ion{Fe}{2} & -0.100 & 84870.912 & VALD \\ 
5534.83 & 5534.839 & \ion{Fe}{2} & -2.900 & 26170.181 & VALD \\ 
        & 5534.894 & \ion{Fe}{2} & -0.440 & 85048.655 & VALD \\ 
5544.46  &  ?      &             &        &           &            \\        	
5549.01 & 5549.000 & \ion{Fe}{2} & -0.190 & 84870.912 & VALD \\ 
5554.89 & 5554.910 & \ion{Fe}{2} & -0.460 & 85680.279 & VALD \\
5558.94 & 5559.05 & \ion{Mn}{2} & -1.310 & 49885.389 & VALD \\ 
5561.30 & 5561.43 & \ion{Mn}{2} & -2.420 & 49893.458 & VALD \\ 
5567.86 & 5567.84 & \ion{Fe}{2} & -1.870 & 54283.234 & VALD \\ 
5570.49  & 5570.54 & \ion{Mn}{2} & -1.430 & 49820.865 & VALD \\ 
5577.96 & 5577.912 & \ion{Fe}{2} & -0.110 & 85462.908 & VALD \\ 
        & 5578.008 & \ion{Fe}{2} & -0.620 & 85495.368 & VALD \\ 
5586.94 & 5586.842 & \ion{Cr}{2} & 0.930 & 88003.158 & VALD \\ 
5588.21 & 5588.221 & \ion{Fe}{2} & 0.160 & 85462.908 & VALD \\ 
5606.14 & 5606.15 & \ion{S}{2} & 0.120 & 110766.562 & VALD \\ 
5645.47 & 5645.39 & \ion{Fe}{2} & 0.190 & 85184.772 & VALD \\
5648.830 & 5648.90 & \ion{Fe}{2} & -0.170 & 85184.772 & VALD \\ 
5651.40 & 5651.52 & \ion{Fe}{2} & -0.610 & 85728.844 & VALD \\ 
5660.15 & 5659.99 & \ion{S}{2} & -0.220 & 110313.403 & VALD \\
5726.59 & 5726.55 & \ion{Fe}{2} & -0.040 & 86416.369 & VALD \\ 
5780.09 & 5780.13 & \ion{Fe}{2} & 0.420 & 86124.348 & VALD \\ 
5783.75 & 5783.62 & \ion{Fe}{2} & 0.370 & 86416.369 & VALD \\
5811.73 & 5811.634 & \ion{Fe}{2} & -0.610 & 86124.348 & VALD \\ 
        & 5811.79  & \ion{Fe}{2} &  -0.640 & 86124.299 &  VALD         \\
	5813.59 & 5813.67 & \ion{Fe}{2} & -2.700 & 44929.533 & VALD \\ 
5842.19 & 5842.29 & \ion{Fe}{2} & -0.330 & 86599.791 & VALD \\ 
5852.44 & 5852.49 & \ion{Ne}{1} & -0.450 & 135888.715 & VALD \\ 
5854.12 & 5854.19 & \ion{Fe}{2} & -0.110 & 86599.791 & VALD \\ 
5871.67  & 5871.80  & \ion{Fe}{2} & -0.640      & 86124.299         &   VALD        \\
5875.75  & 5875.599  & \ion{He}{1} &-1.510 & 169086.769 &  hfs  \\
         & 5875.614  & \ion{He}{1} & -0.340 & 169086.769 &       \\
	 & 5875.615  & \ion{He}{1} & 0.410 & 169086.769 &          \\
	 & 5875.615  & \ion{He}{1} & -0.340 & 169086.845 &         \\
	 & 5875.640  & \ion{He}{1} & 0.140  & 169086.845 &         \\
	 & 5875.966  & \ion{He}{1} & -0.210 & 169087.834 &          \\ 
5902.95  & 5902.83  & \ion{Fe}{2} & 0.420   & 86416.331 & VALD           \\
5961.68  & 5961.71  & \ion{Fe}{2} & 0.700   & 86124.299 &  VALD           \\
5978.85  & 5978.93  & \ion{Si}{2} & 0.000   & 81251.320 &  VALD            \\ 
6096.19 & 6096.16 & \ion{Ne}{1} & -0.300 & 134459.290 & VALD \\ 
6122.48 & 6122.45 & \ion{Mn}{2} & 0.950 & 82136.483 & VALD \\ 
6143.04 & 6143.06 & \ion{Ne}{1} & -0.100 & 134041.838 & VALD \\ 
6147.65 & 6147.74 & \ion{Fe}{2} & -2.800 & 31364.454 & VALD \\ 
6149.28 & 6149.26 & \ion{Fe}{2} & -2.800 & 31368.453 & VALD \\ 
6156.57& 6155.961 & \ion{O}{1} & -1.360 & 86625.757 & hfs \\ 
        & 6155.980 & \ion{O}{1} & -1.010 & 86625.757 &  \\ 
          & 6155.989 & \ion{O}{1} & -1.120 & 86625.757 &  \\ 
          & 6156.737 & \ion{O}{1} & -1.490 & 86627.777 &  \\ 
          & 6156.755 & \ion{O}{1} & -0.900 & 86627.777 & \\ 
         & 6156.770 & \ion{O}{1} & -0.690 & 86627.777 &  \\ 
6158.14 & 6158.172 & \ion{O}{1} & -1.000 & 86631.453 & hfs \\ 
        & 6158.180 & \ion{O}{1} & -0.410 & 89511.404 &  \\ 
6163.69   & 6163.59  & \ion{Fe}{2} & -3.170       & 50157.455  &  VALD          \\
6175.13 & 6175.14 & \ion{Fe}{2} & -2.090 & 50187.824 & VALD \\ 
6238.32 & 6238.39 & \ion{Fe}{2} & -2.800 & 31364.454 & VALD \\ 
6239.65 & 6239.610 & \ion{Si}{2} & 0.180 & 103556.025 & VALD \\ 
        & 6239.660 & \ion{Si}{2} & 0.020 & 103556.156 & VALD \\ 
6247.47 & 6247.56 & \ion{Fe}{2} & -2.400 & 31387.979 & VALD \\ 
6334.33 & 6334.43 & \ion{Ne}{1} & -0.310 & 134041.838 & VALD \\ 
6347.00 & 6346.737 & \ion{Mg}{2} & 0.030 & 93310.590 & VALD \\ 
         & 6346.962 & \ion{Mg}{2} & -0.130 & 93311.112 & VALD \\ 
          & 6347.100 & \ion{Si}{2} & 0.150 & 65500.472 & VALD \\ 
6357.19 & 6357.16 & \ion{Fe}{2} & 0.240 & 87985.668 & VALD \\ 
6371.23 & 6371.36 & \ion{Si}{2} & -0.080 & 65500.472 & VALD \\ 
6383.22  &   ?    &             &        &           &            \\  
6402.19 & 6402.25 & \ion{Ne}{1} & 0.350 & 134041.838 & VALD \\ 
6416.84 & 6416.92 & \ion{Fe}{2} & -2.900 & 31387.979 & VALD \\ 
6432.67 & 6432.68 & \ion{Fe}{2} & -3.500 & 23317.635 & VALD \\ 
        & 6432.68 & \ion{Fe}{2} & -3.710 & 23317.632 &  VALD         \\
6456.22 & 6456.38 & \ion{Fe}{2} & -2.200 & 31483.198 & VALD \\ 
6506.56  & 6506.53  & \ion{Ne}{1} & 0.030 & 134459.290 &  VALD         \\
6577.99   & 6578.05  & \ion{C}{2}  & 0.120 & 116537.649 &  VALD         \\
6582.92   & 6582.88  & \ion{C}{2}  & -0.180 &  116537.649  & VALD           \\ 
6678.10 & 6678.15 & \ion{He}{1} & 0.329 & 171134.897 & VALD \\ 
\hline 
\end{longtable}
\end{center}

\end{document}